\documentclass[aps,prl,twocolumn,showpacs,groupedaddress,letterpaper]{revtex4-1}  
\usepackage{graphicx}  
\usepackage{subfigure}

\newcommand{\beq}{\begin{equation}}
\newcommand{\eeq}{\end{equation}}
\newcommand{\bea}{\begin{eqnarray}}
\newcommand{\eea}{\end{eqnarray}}

\begin{document}

\title{Reentrant behavior of divalent counterion mediated DNA-DNA electrostatic interaction}

\author{SeIl Lee, Tung T. Le, and Toan T. Nguyen}

\affiliation{School of Physics, Georgia Institute of Technology, 
837 State Street, Atlanta, Georgia 30332-0430}


\begin{abstract}
The problem of DNA-DNA interaction mediated by divalent counterions 
is studied using computer simulation. Although divalent counterions
cannot condense free DNA molecules in solution, we show that if DNA configurational
entropy is restricted, divalent counterions can cause DNA reentrant
condensation similar to that caused by tri- or tetra-valent counterions.
DNA-DNA interaction is strongly repulsive at small or large counterion
concentration and is negligible or slightly attractive for a concentration in between. 
Implications of our results to experiments of DNA ejection from bacteriophages
are discussed.
The quantitative result serves to understand electrostatic effects
in other experiments involving DNA and divalent counterions.
\end{abstract}

\pacs{87.19.xb, 87.14.gk, 87.16.A-}

\maketitle

The problem of DNA condensation
has seen a strong revival of interest in recent years 
because of the need to develop effective ways of
gene delivery for the growing field of genetic therapy. 
DNA viruses such as bacteriophages
provide excellent study candidates for this purpose. 
One can package genomic DNA into viruses, 
then deliver and release the molecule into targeted individual cells. 
Recently there is a large biophysic literature dedicated to the problem of 
DNA condensation (packaging and ejection) inside bacteriophages 
\cite{*[{For a review, see }]GelbartVirusReview2009}.

Because DNA is a strongly charged molecule in aqueous solution, the process of
ejection of DNA from bacteriophages can be strongly influenced by the 
screening condition of the solution. By varying the salinity of solution,
 one can vary the amount of DNA ejected. Interestingly, monovalent
counterions such as Na$^{+}$ have negligible effect on 
the DNA ejection process \cite{Gelbart03}.
In contrast, multivalent counterions such as Mg$^{+2}$, CoHex$^{+3}$,
Spd$^{+3}$, or Spm$^{+4}$ exert
strong and non-monotonic effects \cite{Knobler08}. There is an optimal counterion
 concentration, $c_{Z,0}$, where the least DNA genome is ejected from the phages. 
For counterion concentration, $c_Z$, higher or lower than $c_{Z,0}$,
more DNA is ejected from phages. The case of divalent counterions is more marginal.
 The non-monotonicity is observed for MgSO$_4$ salt but not for MgCl$_2$ salt up to 
the concentration of 100mM. 

The problem of DNA condensation by divalent counterions is a complex problem
due to contributions from many physical factors. In the literature, most of
the studies dealing with this problem have focused on the ion-specific effects.
For example, the hydration effects have been 
proposed to explain the above dependence on the type of divalent salts 
\cite{Knobler08}. In this paper, we focus on role of non-specific
electrostatic interactions between DNA and counterions.
In a recent work \cite{NguyenPRL2009}, we suggested that some aspects of 
DNA ejection in the presence of divalent counterions can be accounted for 
from the electrostatic point of view.
Specially the strong, non-monotonic influence of divalent counterions
on DNA ejection mentioned above is expected to have the same 
physical origin as the phenomenon of reentrant DNA condensation in free solution 
\cite{NguyenJCP2000,SaminathanBiochem1999,*LivolantBJ1996}.
The fact that divalent counterions can have such strong
influence on DNA ejection is not trivial.
Unlike counterion of higher valences,
Mg$^{+2}$ counterions are known to not condense
DNA \cite{Parsegian92}, or to condense them 
only partially in free solution \cite{Hud01}. However, DNA viruses
provide a unique experimental setup. The constraint of
the viral capsid strongly eliminates configurational entropic cost of
packaging DNA. This allows divalent counterions to influence
 DNA condensation similar to that of tri- or tetra-valent
counterions. 
In this paper, we use computer simulations to study the problem of 
DNA condensation in the presence of divalent counterions.  
We show that indeed, if one includes only the non-specific 
electrostatic contribution,
divalent counterions can induce DNA reentrant 
condensation like those observed for higher counterion valences.
We offer an explanation for the discrepancy between DNA condensation
in free solution versus DNA condensation inside viruses.
Our results show that, in addition to ion-specific
effects, electrostatics exert a strong, non-negligible influence
on qualitative and quantitative behaviors of this system.
The results presented here can provide understanding of not only the
electrostatics of DNA ejection problem, but also can serve as a starting point for 
investigating other systems involving DNA and divalent counterions
where the physical pictures are still not very well understood.

%
\begin{figure}[ht]
\resizebox{6cm}{!}{\includegraphics{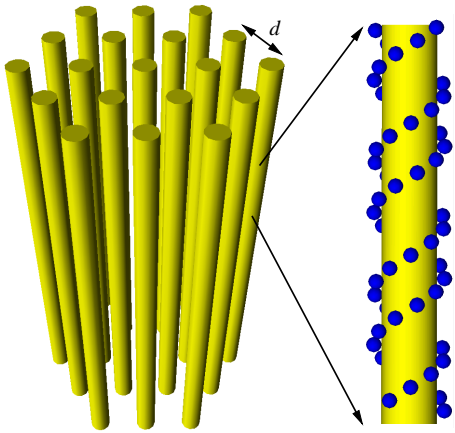}}
\caption{(Color online) A DNA bundle is modelled as a hexagonal lattice
with lattice constant $d$. An Individual
DNA molecule is modeled as a hard-core cylinder with negative charges
glued onto it according to the positions of nucleotides of a B$-$DNA structure.
}
\label{fig:DNA}
\end{figure}
We model the DNA bundle in hexagonal packing as a number of DNA molecules 
arranged in parallel along the $Z$-axis. In the horizontal plane, the DNA 
molecules form a two dimensional hexagonal lattice with lattice constant
$d$ (the DNA$-$DNA interaxial distance) (Fig. \ref{fig:DNA}). 
An individual DNA molecule
is modeled as an impenetrable cylinder with negative charges glued onto it. 
The charges are positioned in accordance with the locations of nucleotide groups
 along the 
double-helix structure of a B$-$DNA. The hardcore cylinder
has radius of 7\AA. The negative charges are hard spheres
of radius 2\AA, charge $-e$, and lie at a 
distance of 9\AA\  from the DNA axis. This
gives an averaged DNA diameter of 1nm. The solvent water is treated as
a dielectric medium with dielectric constant $\varepsilon = 78$
and temperature $T=300^oK$. The dielectric constant mismatch between
 water and DNA interior is neglected, and the cylinder 
only acts to prevent ion penetration. 
In our simulation, the positions of DNA molecules are fixed in space. 
This mimics the constraint on DNA configurational entropy inside viruses and 
other experiments of DNA condensation using divalent counterions. 

In the experiment of DNA ejection from bacteriophages, there are both
monovalent and divalent salts in solution. At very low concentration of 
divalent counterions, DNA is screened mostly by monovalent counterions. To
account for this limit, we include both salts in our simulations.
The mobile ions are modeled as hard spheres with 
unscreened Coulomb interaction (the primitive ion model). The radii of the
coions and monovalent counterions are set to 2\AA. 
(For simplicity, we assume the two salts have the same coions.)
The divalent counterions radius is set to 2.5\AA. 
The interaction between two ions,
$i$ and $j$, with radii, $\sigma_{i,j}$, and charges, $q_{i,j}$, is given by
\beq
U = \left\{
  \begin{array}{l l}
    q_iq_j/\varepsilon r_{ij}& \quad \mbox{if $r_{ij} > \sigma_i + \sigma_j$}\\
    \infty & \quad \mbox{if $r_{ij} < \sigma_i + \sigma_j$}\\ 
  \end{array}
    \right.
\eeq
where $r_{ij}=|\mathbf{r}_i-\mathbf{r}_j|$ is the distance
between the ions. 

The simulation is carried out using the periodic boundary condition. 
A periodic simulation cell with 
$N = 12$ DNA molecules in the horizontal $(x,y)$ plane and 3 full helix 
periods in the $z$ direction is used. The dimensions of the box are $L_x = 3d$,
 $L_y=2\sqrt{3}d$, and $L_z = 102$\AA. The long-range electrostatic interactions
between charges in neighboring cells are treated using the
Ewald summation method. 
In Ref. \cite{Nordenskiold95,*NordenskioldJCP86}, it is shown that
the macroscopic limit is reached when $N \geq 7$. Our simulation cell
contains 12 DNA helices,
hence it has enough DNA molecules to eliminate the finite size effect.
We did test runs with 1, 4, 7, and 12 DNA molecules to verify that
this is indeed the case. They are also used
to check the correctness of our computer program by reproducing the results
of DNA systems studied in Ref. \cite{Nordenskiold95,*NordenskioldJCP86}
in specific limits.

In a practical situation, the DNA bundle is in equilibrium
with a water solution containing free mobile ions at a given concentration.
Therefore we simulate the system using Grand Canonical Monte-Carlo (GCMC) simulation.
The number of ions are not constant during the simulation. Instead
their chemical potentials are fixed. The chemical potentials
are chosen in advance by simulating a DNA$-$free salt solution
and adjusting them so that the solution has the correct
ion concentrations. In a simulation, the ions are
inserted into or removed from the system in groups to maintain
the charge neutrality \cite{CohenGCMC}. 
Following Ref. \cite{CohenGCMC}, instead of using 
individual chemical
potentials, $\mu_{+2}$, $\mu_{+1}$, and $\mu_{-1}$, for each ion species,
we use only the combined chemical potentials, 
\beq
\mu_{+2}^{salt} = \mu_{+2} + 2\mu_{-1}, ~~
\mu_{+1}^{salt} = \mu_{+1} + \mu_{-1},
\eeq
in the Metropolis acceptance criteria of a particle insertion/deletion move.
In this paper, we simulate DNA bundles at varying concentrations $c_Z$. 
%
Both $\mu_{+1}^{salt}$ and $\mu_{+2}^{salt}$ are
adjusted so that the monovalent salt bulk concentration, $c_1$, in the DNA$-$free 
solution is always at 50mM (typical value of the DNA ejection experiment)
and $c_Z$ is at the desired value. Typical standard deviations in the final salt 
concentrations are about 10\%.


To study DNA$-$DNA interactions, we use the Expanded Ensemble
method \cite{Nordenskiold95} to calculate the pressure
of the DNA bundle. In this method, we calculate the difference 
of the system free energy at different volumes by sampling 
these volumes simultaneously in a simulation run.
By calculating the free energy difference $\Delta \Omega$
for two nearly equal volumes, $V$ and $V+\Delta V$, we
can calculate the total pressure of the system,
$
P(T,V,\{\mu_\nu\}) 
    \simeq -\Delta \Omega/\Delta V
$
(here $\{\mu_\nu\}$ is the set of chemical potentials
of different ion species). The osmotic pressure of the DNA bundle is 
then obtained by subtracting the total pressure of the bulk DNA$-$free 
solution, $P_b(T,V,\{\mu_\nu\})$, from the total pressure of the DNA system,
$
P_{osm} (T,V, \{\mu_\nu\}) = P(T,V, \{\mu_\nu\}) - P_b(T,V,\{\mu_\nu\})
$.

\begin{figure}[ht]
\subfigure[][]{\resizebox{7cm}{!}{\includegraphics{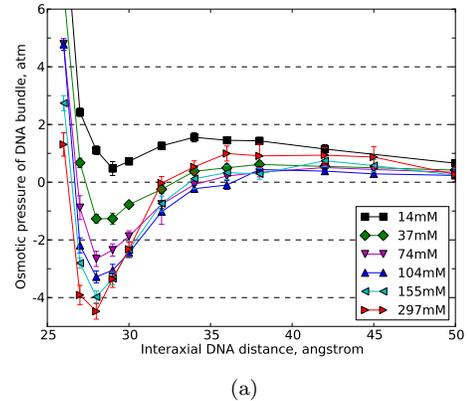}}}
\subfigure[][]{\resizebox{7cm}{!}{\includegraphics{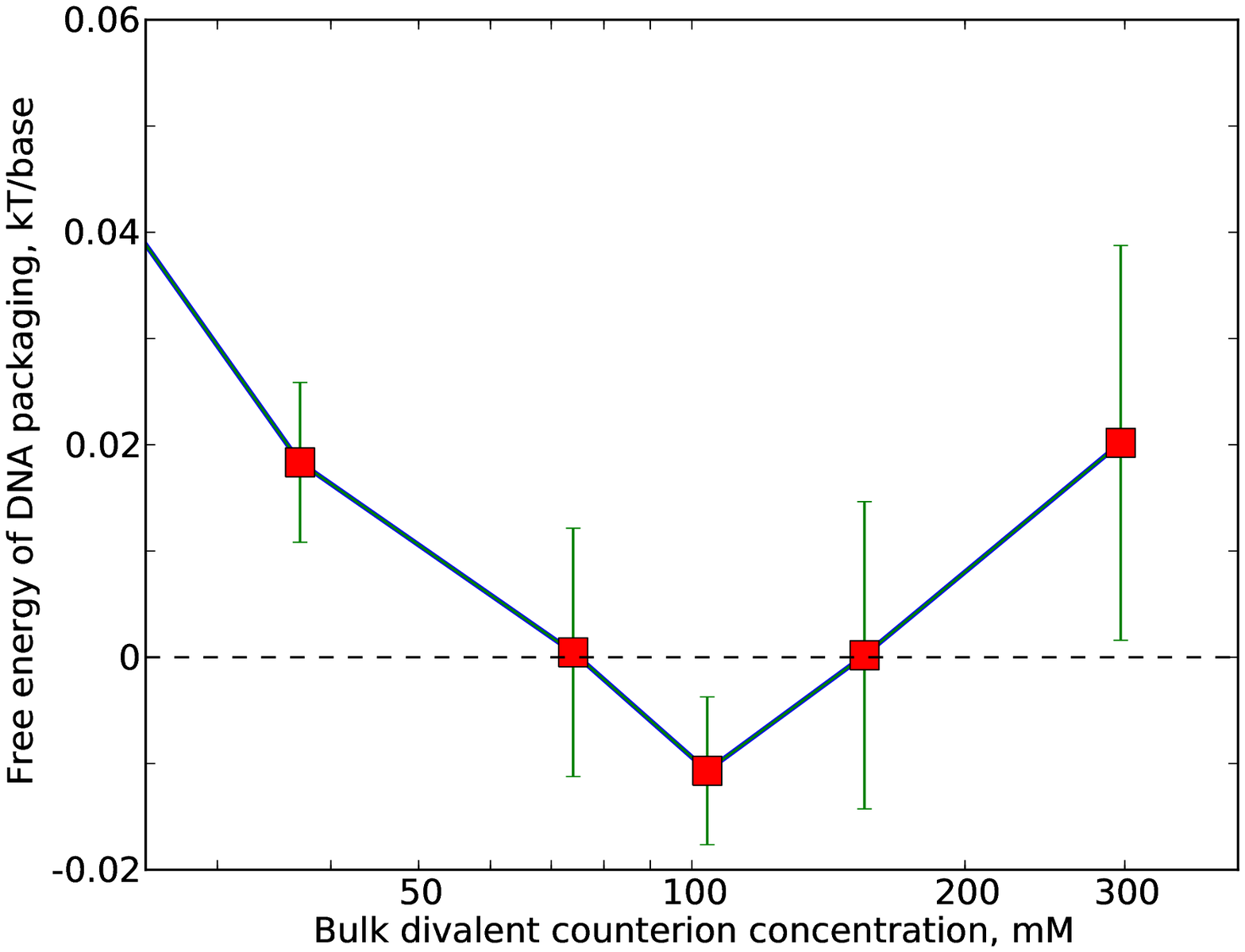}}}
\caption{(Color online) a) The osmotic pressure of the DNA bundle
as a function of the interaxial DNA distance $d$ for different
divalent counterion concentration $c_Z$ shown in the inset. 
b) The numerically calculated
free energy of packaging DNA molecules
into hexagonal bundles as a
function of the divalent counterion concentrations.
}
\label{fig:posm}
\end{figure}

In Fig. \ref{fig:posm}a, the osmotic pressure of the DNA bundle
at different $c_Z$ is plotted as a 
function of the interaxial DNA distance, $d$.
Because this osmotic pressure is directly related to
the ``effective'' force between DNA molecules at that interaxial
distance \cite{Nordenskiold95,*NordenskioldJCP86}, 
Fig. \ref{fig:posm}a also serves as a plot of DNA$-$DNA interaction. 
As one can see, when $c_Z$ is greater than a value 
around 20mM, there is a short$-$range attraction between two DNA molecules as 
they approach each other. This is the well-known phenomenon of like-charge attraction
between macroions \cite{NetzLikeChargedRods,*GelbartPhysToday, NguyenJCP2000}. 
It is the result of the electrostatic
correlations between counterions condensed on the surface of
each DNA molecule. The attraction appears when the distance between these
surfaces is on the order of the lateral separation
between counterions (about 14\AA\ for divalent counterions). The maximal attraction 
occurs at the distance $d\simeq 28$\AA\ in good agreement with various theoretical and 
experimental results \cite{Parsegian92,Phillips05}. 
For smaller $d$, the DNA-DNA interaction
experiences a sharp increase due to the hardcore repulsion
between the counterions. One also sees that the depth of
attractive force between DNA molecules saturates at around $-4$ atm as $c_Z$ increases. 
This saturation is easily understood. At small $c_Z$, there are both 
monovalent and divalent counterions present in the bundle.
As $c_Z$ increases, divalent counterions replace
monovalent ones in the bundle as the later ions
are released into the bulk solution to increase the overall 
entropy of the solution. However, charge neutrality condition of the DNA macroscopic
bundle and the hardcore repulsion between ions limit
how many divalent counterions can be present inside the bundle. 
Once all monovalent counterions are released into solution
(replaced by divalent counterions), further
increase in $c_Z$ does not significantly change
the number of divalent counterions in the bundle. This leads to the
observed saturation of DNA$-$DNA short$-$range attraction with increasing $c_Z$.

The strong influence of divalent counterions on DNA bundles
can be seen by looking at the DNA-DNA ``effective" interaction
at larger $d$. 
As evident from Fig. \ref{fig:posm}a for large $d$, at small $c_Z$
DNA-DNA interaction is repulsive. As $c_Z$ increases,
DNA-DNA interaction becomes less repulsive and reaches a minimum around 
100mM. As $c_Z$ increases further, DNA-DNA repulsion starts 
to increase again. 
This non-monotonic dependence of DNA-DNA ``effective" interaction
on the counterion concentration is even more clear if one
calculates the free energy of packaging DNA into bundles. This free 
energy is the difference between the free energy of 
a DNA molecule in a bundle and that of an  individual DNA molecule
in the bulk solution.
Per DNA nucleotide base, this free energy is given by:
\beq
\mu_{\mbox{DNA}}(d) = (l/L_z N)
  \int_\infty^d P_{osm}(d') dV(d') ,
\eeq
%
%
here $l=1.7$\AA\  is the distance between DNA nucleotides along the
axis of the DNA, and $V(d)=L_xL_yL_z=6\sqrt{3}L_zd^2$ is the volume of
our simulation box. The result for $\mu_{\mbox{DNA}}(d)$ at
the optimal bundle lattice constant $d=28$\AA\  is plotted in Fig. 
\ref{fig:posm}b as function of the $c_Z$ \cite{NIntFootnote}.
Once again, 
there is an optimal concentration, $c_{Z,0}\simeq 100mM$, where the
free energy cost of packaging DNA is lowest. 
It is even negative indicating the tendency of the divalent
counterions to condense the DNA. At smaller or larger concentrations
of the counterions, the free energy cost of DNA packaging
is higher.

\begin{figure}[ht]
\resizebox{7.5cm}{!}{\includegraphics{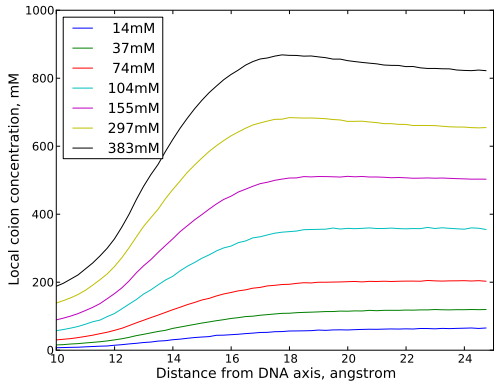}}
\caption{(Color online) Local concentration of coions as a function of
distance from the axis of a DNA in the bundle for different
divalent counterion concentrations, for $d=50$\AA.
}
\label{fig:gr_coion}
\end{figure}
This reentrant behavior of DNA interaction can be understood.
At large separations, the distribution of counterions
in the bundle can be considered to be composed of two populations: 
condensed layers of counterions near the surfaces of the
DNA molecules and diffuse layers of counterions further away.
It is reasonable to expect the thickness of the condensed counterion layer
to be on the order of
the average lateral distance between counterions on the DNA surface ($\approx 14$\AA).
So for $d \geq 34$\AA, both counterion populations are present and one expects DNA-DNA 
interaction to be the standard screened Coulomb interaction
between two charged cylinders with charge density $\eta^*$.
The qualitative dependence of $\eta^*$ on $c_Z$ can be obtained by plotting 
the local coion concentration $c_{-1}$ as a function of distance
from DNA axis (Fig. \ref{fig:gr_coion}).
At low $c_Z$, $c_{-1}$ decreases as
$d$ decreases from $\infty$ suggesting $\eta^*$ is negative (undercharged DNA). 
At high $c_Z$, $c_{-1}$ increases
as $d$ decreases until the condensed counterion layers start to overlap
at $d \approx 34$\AA. This
shows that $\eta^*$ is positive (overcharged DNA). 
In both cases, DNA repulsion is strong. For an intermediate value of $c_Z$, 
$\eta^* \simeq 0$, DNA is almost neutral, and the repulsion is weakest. 
Furthermore, the like-charge attraction
 among DNA molecules mediated by the counterions \cite{NetzLikeChargedRods} 
is dominant in this concentration range, causing the electrostatic packaging
 free energy to become negative.


%
Figure \ref{fig:posm}b gives 
a value of $-0.001k_BT$/base for the short$-$range attraction among DNA 
molecules at the optimal concentration.
This is slightly less negative than previous theoretical fit of viral DNA ejection
 experiments \cite{NguyenPRL2009}.
We believe this small difference is due our choice of the 
system's physical parameters such as ion sizes
\cite{BruinsmaPRLRod,*NordenskioldPRL98,*DNA_reentrantPRE}. 
The azimuthal orientation correlations of DNA \cite{KornyshevRMP2007}
are another omission in our study. Relaxation along this degree
of freedom can further lowering energy of the system.
The non-electrostatic (such as van der Waals) interactions at small $d$
can also enhance DNA attraction. On the other hand, dielectric constant mismatch
between water and DNA interior could push the condensed ions
away from DNA interior and lower the attraction energy.
However, these effects are minor at large DNA$-$DNA separations,
thus do not change the qualitative reentrant condensation picture. More
comprehensive studies that take these effects into account are
the subjects of our future works. Nevertheless, the value range of
$-0.001 k_BT$ obtained in this paper is significant.
 It explains why divalent counterions exert
strong effect on DNA ejection from virus but are not able to
condense DNA in free solution. This value corresponds to an attraction
of a fraction of $-k_BT$ per one persistence length ($\simeq 300$ bases). This
is too small to overcome thermal fluctuation of DNA 
(about one $k_BT$ per persistence length), 
thus cannot condense them. Only inside the confinement of the viral capsid,
where DNA configuration entropy is strongly suppressed, 
can divalent counterions cause strong influence. 
The non-monotonic behavior described above has the same
physics as the phenomenon of reentrant DNA condensation by 
counterions \cite{NguyenJCP2000,Shklovskii1999,*NguyenRMP2002}
of high valences. In this paper, we demonstrate clearly that
it can happen to divalent counterions if DNA configuration
entropy is restricted. This correlates well with experimental data of
DNA ejection from bacteriophages. We should mention here that
DNA condensation by divalent counterions has
also been observed in another environment where DNA configuration is 
constrained, namely the condensation of DNA in two dimensional 
systems \cite{Koltover2000}. This fact once again strongly supports
our argument.

In conclusion, in this paper, we use a computer simulation to study 
the electrostatics of DNA condensation in the presence
of divalent counterions. The entropy of DNA configure fluctuation
is suppressed in simulation. Such study can be applied directly
to the experimental problem of DNA ejection from bacteriophages where
DNA condensed in a strongly confined environment. 
Our results show that, even at the level of non-specific electrostatic
interaction, divalent counterions can strongly influence DNA ejection.
This potentially opens up an additional degree of freedom in controlling
bacteriophages for the general purpose of gene therapy or viral
diseases treatment. Beyond the scope of DNA ejection experiments,
we believe the quantitative results of our paper can be used 
to understand many other
experiments involving DNA and divalent counterions.

\begin{acknowledgments}
We would like to thank Lyubartsev, Nordenski\"{o}ld, Shklovskii, Evilevitch,
Fang, Gelbart, Phillips, Podgornick, Rau, and Parsegian for valuable discussions. 
TN acknowledges the support of junior faculty from the Georgia Institute of Technology.
SL acknowledges financial support from Korean-American Scientists
and Engineers Association (Georgia chapter). The authors are indebted
to Dr. Lyubartsev for providing us with the source code of their simulation
program. This code forms the basis of the simulation program
used in this work. TN acknowledges the hospitality of the Aspen Center for
Physics and the Fine Theoretical Physics Institute where part of the work is done.
\end{acknowledgments}

\bibliography{DNA_reentrant}

\begin{thebibliography}{23}
\bibitem [{NIn()}]{NIntFootnote}%
  \BibitemOpen
  \href@noop {} {}\bibinfo {note} {Due to the limitation of computer
  simulations, the numerical integration is performed from $d=28$\AA\ to
  $d=50$\AA\ only. The omitted integration from $d=50$\AA\ to $d=\infty$ gives
  an almost constant shift to $\mu_{\mbox{DNA}}$, and do not change the
  conclusion of this paper.}\BibitemShut {Stop}%
\end{thebibliography}

\end{document}